# Machine Learning Enables Optimization of Diamond for Quantum Applications

Dane W. deQuilettes,[1,*] Eden Price,[1*] Linh M. Pham,[1] Arthur Kurlej,[1] Swaroop Vattam,[1] Alexander Melville,[1] Tom Osadchy,[1] Boning Li,[2] Guoqing Wang,[2,3] Collin N. Muniz,[1] Paola Cappellaro,[2,3] Jennifer M. Schloss,[1] Justin L. Mallek,[1] Danielle A. Braje[1]

1. Massachusetts Institute of Technology, Lincoln Laboratory, Lexington, MA, 02421, USA.

2. Department of Physics, Massachusetts Institute of Technology, Cambridge, MA, 02139, USA.

3. Department of Nuclear Science and Engineering, Massachusetts Institute of Technology, Cambridge, MA, 02139, USA.

**\*Equal Contribution**

**Corresponding Author:** braje@ll.mit.edu

DISTRIBUTION STATEMENT A. Approved for public release. Distribution is unlimited. AFRL-2025-4999
This material is based upon work supported by the Department of the Air Force under Air Force Contract No. FA8702-15-D-0001 or FA8702-25-D-B002. Any opinions, findings, conclusions or recommendations expressed in this material are those of the author(s) and do not necessarily reflect the views of the Department of the Air Force. © 2025 Massachusetts Institute of Technology.

**Abstract:**

Spins in solid-state materials, molecules, and other chemical systems have the potential to impact the fields of quantum sensing, communication, simulation, and computing. In particular, color centers in diamond, such as negatively charged nitrogen vacancy (NV⁻) and silicon vacancy centers (SiV⁻), are emerging as quantum platforms poised for transition to commercial devices. A key enabler stems from the semiconductor-like platform that can be tailored at the time of growth. The large growth parameter space makes it challenging to use intuition to optimize growth conditions for quantum performance. In this paper, we use supervised machine learning to train regression models using different synthesis parameters in over 100 quantum diamond samples. We train models to optimize NV⁻ defects in diamond for high sensitivity magnetometry. Importantly, we utilize a magnetic-field sensitivity figure of merit (FOM) for NV magnetometry and use Bayesian optimization to identify critical growth parameters that lead to a 300% improvement over an average sample and a 55% improvement over the previous champion sample. Furthermore, using Shapley importance rankings, we gain new physical insights into the most impactful growth and post-processing parameters, namely electron irradiation dose, diamond seed depth relative to the plasma, seed miscut angle, and reactor nitrogen concentration. As various quantum devices can have significantly different material requirements, advanced growth techniques such as plasma-enhanced chemical vapor deposition (PE-CVD) can provide the ability to tailor material development specifically for quantum applications.

**Introduction:**

Solid-state materials have emerged as an exciting platform for quantum technologies. Their tunable material properties can be adapted to scalable manufacturing processes, while the simplicity of the crystal lattice enables vacuum-free devices that are robust to a range of conditions such as varying temperature, pressure, and pH levels.[1,2] Materials chemistry offers the ability to tune quantum properties and optimize performance for specific quantum applications.[1] Tailoring quantum properties often requires full control over the local chemical environment, surfaces and interfaces, and bulk material quality.[3] Optimizing this complex environment is hindered by a slow feedback loop between synthesis and characterization, and device testing often requires specialized expertise. We shorten this loop by targeting growth with specific material processing parameters relevant to device performance.

As an example, we show a specific case of optimizing growth for magnetometry applications. A central focus of NV centers is their utilization as quantum sensors, where they can be used to detect physical quantities such as temperature, pressure, electric, and magnetic fields over a large dynamic range with vector capabilities.[4,5] Key metrics for quantifying sensor performance are the minimum detectable change as well as the limit of detection of the physical quantity. For diamond-based magnetometry applications, a magnetic field sensitivity at <1 $pT/\sqrt{Hz}$ in the ~Hz to GHz region was recently demonstrated,[6] with some sensing protocols promising further reductions down to 1 $fT/\sqrt{Hz}$.[7] If achieved, this could enable cheap, portable solutions for magnetoencephalography and magnetocardiography that are currently dominated by more sensitive superconducting quantum interference devices (SQUIDs) and vapor-cell-based magnetometers.[8] Importantly, these types of advancements require improvements in quantum materials that are tailored to achieve optimal performance in addition to advanced noise suppression techniques.

For NV diamond color centers, optimization can be achieved by enhancing properties such as coherence times, NV⁻/N conversion efficiency, charge state stability (i.e. keeping NV⁻ rather than converting to charge neutral NV$^0$), and reducing crystal strain.[9] Recently, some studies note dependence of coherence times on growth conditions,[10,11] but these data sets are limited to a small section of process parameter space. The complex parameter space for diamond growth and post-processing makes it time and cost intensive to globally optimize. Data-informed materials development has the ability to accelerate the optimization of diamond for a wide-range of quantum applications, such as optically detected magnetic resonance (ODMR) NV-based magnetometers.[12] In this work, we demonstrate this cycle by using machine learning to pair quantum materials synthesis and characterization feedback.

We specifically focused on quantum magnetometry as there is a clear relationship between measurable optical and quantum diamond properties and the resulting quantum sensor performance.[13] Therefore, we characterize key photophysical properties which are combined into a single figure of merit (See Eq. 1) to compare different growth recipes and reveal how processing parameters impact material quality. The previously derived Ramsey single-shot magnetic field sensitivity for ensembles serves as our sensitivity figure of merit (FOM). [9,11,14]

$$FOM \sim \frac{1}{g_e \mu_B} \frac{1}{C} \frac{1}{\sqrt{\beta n_{NV-} V}} \frac{\sqrt{T_{dead} + T_2^*}}{T_2^*} \qquad \text{(Eq. 1)}$$

In Equation 1, $g_e$ is the electron g-factor, $\mu_B$ is the Bohr magneton, $C$ is the measurement contrast, $\beta$ is the optical collection efficiency, $V$ is the excitation volume, $n_{NV-}$ is the NV density, $T_{dead}$ is the dead time for initialization and readout, and $T_2^*$ is the Ramsey coherence time.

Importantly, for comparing quantum diamond, this figure of merit captures the tradeoff between higher NV densities leading to shorter coherence times.[9,11] We also use it to quantify the impact of laser power density on magnetometer sensitivity, which is often overlooked.[11] In particular, laser power density can determine the charge-state ratio and overall performance of the sensor depending on the measurement protocol (i.e. ODMR, Ramsey, dynamical decoupling schemes, etc.). In addition, this FOM can be used for both DC and AC sensing, by changing the coherence time definition from T$_2$* to T$_2$, as well as for single NV's where $n_{NV-} = 1$.[9]

**Diamond Growth/Processing Inputs for Training ML Regression Models**

A first step in applying machine learning is to capture the material processing parameters that will most likely have an impact on the sensitivity FOM defined above. In Figure 1, we detail the five-step process of creating the data set used for optimization with ML, including 1) measuring the high-pressure, high-temperature (HPHT) diamond seed properties 2) tracking the PE-CVD growth parameters 3) performing various electron irradiation and annealing protocols to improve defect concentrations and properties and 4) characterizing optical and quantum properties of the N-doped diamond. Importantly, all of this data is combined into a custom database where the sensitivity FOM is calculated and used for the last step: 5) training ML regression models. Finally, the ML models are paired with a global optimization method, known as Bayesian optimization (BO), to suggest new growth conditions. We then use these suggestions to grow new samples and further refine the model, a process called active learning.

Figure 1 shows example properties recorded in a custom database over four main steps including seed, growth, and post-growth processing, as well as characterization of NV properties. Seed properties include parameters such as miscut angles $\theta$ and $\kappa$ as well as thickness ($t$), where $\theta$ is the angle relative to the vector normal to the (100) surface and $\kappa$ is the angle from the (001) plane (see SI for details). The crystallographic miscut angle $\theta$ of the seed is included as it has been shown to impact the homoepitaxial growth rate and defect incorporation.[15,16,17,18] For growth, we include the chamber pressure, gas composition, growth time, and distance of the seed from the plasma ball (i.e., seed depth), which all impact the growth kinetics and surface morphology (step flow versus island).[19,20,21] After diamond growth, the material is typically processed through electron irradiation and annealing to optimize NV⁻ formation, coherence time, and charge state stability.[22] In all, we track over 100 diamonds each with more than 10 unique processing parameters. Such a large, multi-dimensional parameter space is challenging to optimize by traditional methods, where previous literature has only explored a single parameter's impact on diamond growth.[15,16,19,23] We study how these variables work in conjunction with each other and impact quantum diamond performance.

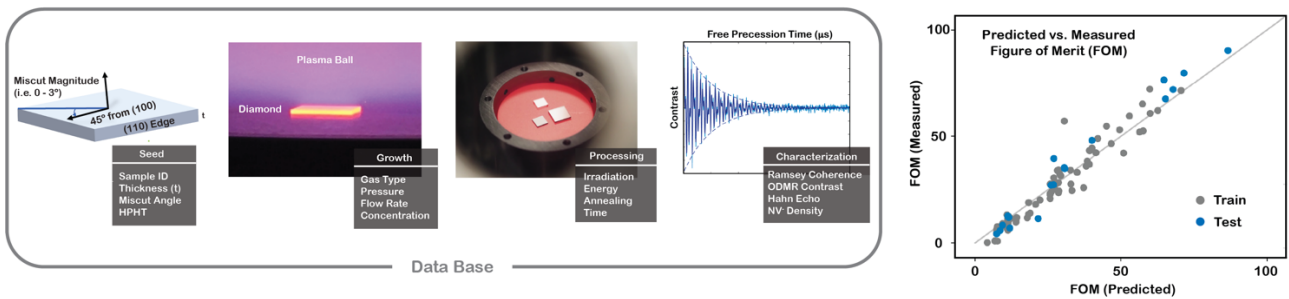

**Figure 1.** Schematic illustration of the workflow for tracking diamond seed, growth, post processing, and NV properties in a custom database. Machine learning is used to optimize a magnetic field sensitivity figure of merit based on material processing parameters. Input processing parameters are listed under each main step and tied to NV optical properties.

**Characterization of Quantum System**

The behavior of NV⁻ defects in diamond are sensitive to their surrounding environment including the presence of charge traps, acceptor states, band bending, other proximal spins, as well as unscreened charges. These sources lead to magnetic and electronic noise as well as charge state instability that can have a large impact on NV⁻ quantum sensor performance. Therefore, we characterize the NV⁻/NV⁰ charge state ratio, coherence time, and total NV⁻ density for each sample. Figure 2 shows an example dataset of these optical and quantum properties for a typical quantum diamond sample. Specifically, Figure 2a shows the NV⁻/NV⁰ as a function of laser power, which is determined by deconvolving the photoluminescence spectra using basis functions.[24] The relative ratio of NV⁻ typically decreases at higher laser powers densities (see Figure 2a insets) due to multiphoton ionization effects.[11,25] In addition, Ramsey and Hahn echo sequences are used to characterize the $T_2^*$ and $T_2$ coherence times of the NV⁻ ensemble, respectively. Figure 3b shows the pulse sequence along with the Ramsey decay envelope of an average electron-irradiated sample, with a $\tau = 3.5 \pm 0.4$ μs. Figure 3c shows the Hahn-echo pulse sequence along with the decay envelope, with a $\tau = 23.2 \pm 1.6$ μs. These values are near average for the dataset and consistent with high quality single crystal diamonds grown with $^{12}$C.[9,11] Finally, we determine the NV⁻ density through calibrated cryogenic ultraviolet visible (UV-Vis) measurements using methods described previously (see Supplementary Information and Supplementary Figure 3).[26]

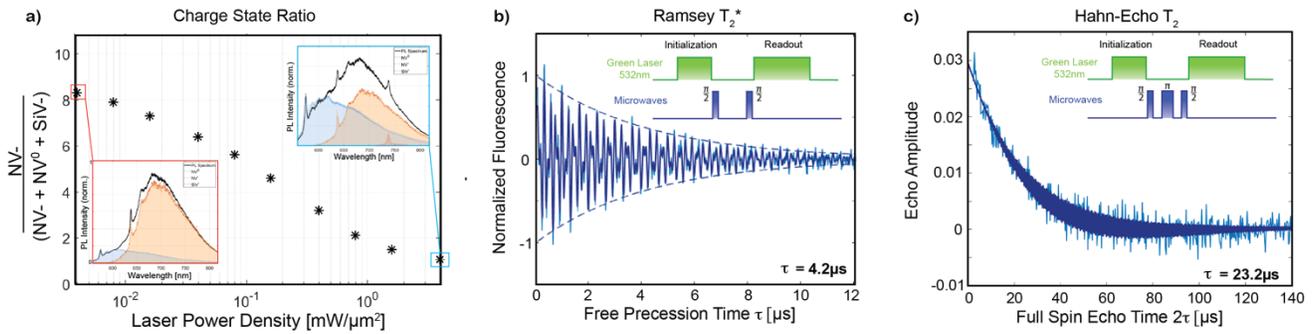

**Figure 2.** a) Fluorescence charge state ratio of negatively charged nitrogen-vacancy (NV⁻) to neutral nitrogen-vacancy (NV⁰) of a typical N-doped diamond plotted versus laser power density ($\lambda$ = 532 nm, continuous wave). Insets show photoluminescence spectra (black solid line) of the diamond at the lowest laser power density (~4 µW/um²) and the highest laser power density (~4 mW/um²), respectively. Spectral decomposition of NV⁰ (blue), NV⁻ (orange), and SiV⁻ (red) are shown in each figure. b) Ramsey $T_2^*$ decay envelope of a typical NV⁻ doped diamond. Inset shows the Ramsey measurement protocol with laser initialization and readout (green) and the applied microwave pulse sequence (blue). c) Hahn-echo decay envelope of a typical NV⁻ doped diamond. Inset shows the Hahn-echo measurement protocol with laser initialization and readout (green) and applied microwave pulse sequence (blue).

**Validation of ML Model and Bayesian Optimization**

From above, we track the various diamond processing inputs (i.e. featurize the data) and NV properties to calculate the quantum sensing FOM. Next, we use supervised machine learning with Bayesian optimization to identify seed, growth, and processing parameters to target higher FOMs and reveal new relationships between process parameters and NV performance.

We train multiple regression models covering linear and quadratic regression, decision trees, and ensemble methods using an 80:20 train-test split.[27] Figure 4a shows the root mean square error (RMSE) of the test data set, with the distributions on each bar reflecting the 5-fold cross-validated standard deviation. We see that gradient boost and random forest regression have the lowest average RMSE values across various train-test splits. To ensure that we are not over-parameterizing the data set, we perform feature down-selection by testing the sensitivity of the RMSE values to feature removal.[28] We find that gradient boost has the best combination of low RMSE and standard deviation and, therefore, use this model in the rest of the study. Figure 4b shows a parity plot of the predicted and measured FOM of the gradient boost model with the model's training data (gray circles) and test data (blue circles). The clustering of the test data points near the parity line shows the high predictive performance of the trained regression model.

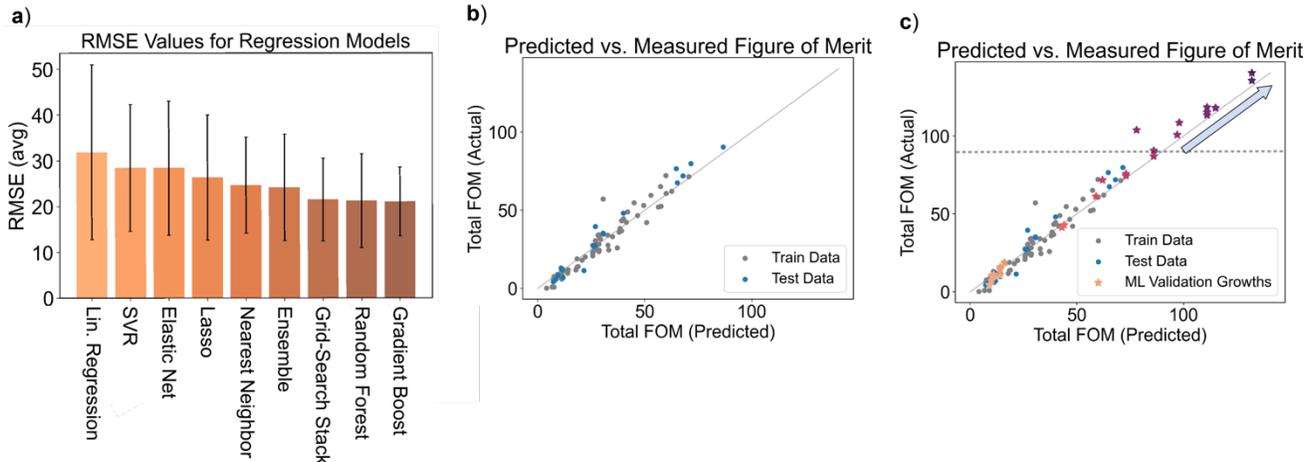

**Figure 3.** a) Average root mean square error (RMSE) across various trained regression models. Each model shows the five-fold cross-validation standard deviation (black whisker plots). b) Parity plot of predicted versus actual sensitivity figure of merit (FOM) values for standardized training data (gray) and test data (blue) for the gradient boost model. c) Parity plot of predicted versus actual FOM (same as 4b) with addition of ten validation growths (stars) determined using Bayesian inference/optimization. Arrow on the graph shows the increase from the highest achieved FOM before ML (dashed gray line) to the highest achieved with ML.

Next, we further test the robustness of the model by performing a set of validation experiments. Specifically, we use a Gaussian process (GP) surrogate function and Bayesian inference/optimization to suggest growth regimes to test the predictive capabilities in unexplored parameter space. In some cases, we weight the acquisition function towards exploration in order to grow in undersampled parameter spaces or towards exploitation to optimize the sensitivity FOM. Figure 4c shows the same training data and test data in Figure 4b, as well as new validation growths obtained from Bayesian optimization. The parity plot contains FOM values calculated from both the pre-irradiated and post-irradiated diamond samples, where the validation samples on the bottom of the graph have lower FOM's (orange stars) due to a lower NV density after growth. Importantly, by tuning the acquisition function towards exploitation, we achieve the highest FOM reported from our lab to date and likely one of the highest quality quantum diamonds ever reported. In only a few months of growing validation samples, the ML-guided optimization results in a 300% FOM improvement over our average sample. We highlight that the ML-guided sample growth also leads to a marked +55% improvement over our previous best sample, which took two years of human-guided, brute force optimization.

In addition to demonstrating a successful optimization strategy, model interpretability is just as important to advance our scientific understanding of the diamond processing space.[29] In this regard, we perform a SHAP (Shapley Additive exPlanations) feature importance ranking to identify which input parameters lead to the largest changes in the sensitivity FOM.[30] Figure 5a shows the SHAP summary plot in order of most to least important processing parameters that impact the FOM along with each parameter's impact on the FOM for every sample. We identify total electron dose as the most important processing parameter, followed by seed depth, miscut angle, and chamber nitrogen concentration. These four processing parameters account for ~75% of the variation in FOM.

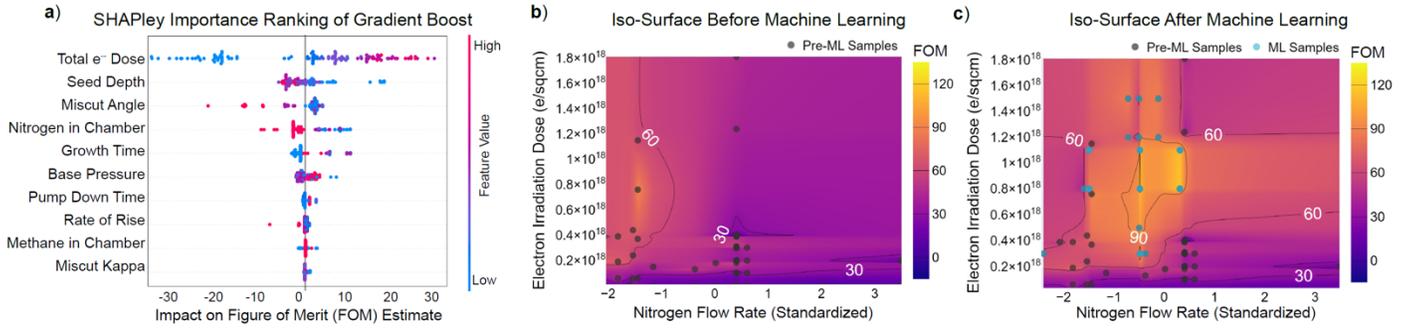

**Figure 4.** a) SHAPley Importance Ranking for the same data set from b) ordered from highest to lowest impact on the FOM. Each growth feature contains the entire data set (N = 97) and is further broken down into the feature value (color scale) and overall impact on the sensitivity FOM estimate. b) 2D iso-surface of the Sensitivity Figure of merit (FOM) (color scale) versus reactor nitrogen concentration (ppm) and electron irradiation dose (e$^-$/cm$^2$), the two highest-impact variables determined from SHAPley Importance Ranking from Gradient Boost using all data (black dots) from before machine learning. c) Updated 2D iso-surface of the Sensitivity FOM (color scale) versus reactor nitrogen concentration (ppm) and electron irradiation dose (e$^-$/cm$^2$) using all data from before machine learning (black dots) as well as validation growths (blue dots).

Although it has previously been shown that irradiation dose has a significant impact on NV$^-$ formation,[25] we quantify the impact of dose relative to other processing parameters. For example, optimizing the dose can change the FOM estimate by as much as 60 (i.e. >50%), with a critical dose threshold of ~ 1x10$^{18}$ cm$^{-2}$, consistent with a previous study.[25] In addition, our SHAP analysis revealed that lower seed depths, which result in higher growth temperatures due to a closer distance between the seed and plasma ball,[31] are critical in optimizing FOM. Beyond this, we identify both lower theta miscut angles and lower reactor nitrogen concentrations critical to achieving higher FOMs. From Equation 1, one might predict that higher chamber nitrogen concentration and doping densities would lead to a higher FOM. However, our results suggest that the greater NV$^-$ signal intensity is overpowered by the reduction in coherence time, which may result from unwanted spin defects forming at higher N-doping densities. Therefore, it is critical to control the seed miscut angle as nitrogen incorporation can be tuned with this parameter.[23]

Finally, to better understand the sensitivity of FOM to the most important SHAP variables, such as irradiation dose and nitrogen doping density, we took cross-sections of the multidimensional FOM surface. Figure 5b shows a cross-section of the FOM as a function of these two variables to generate an isosurface based on the training and test data set (black circles). In Figure 5c, we regenerate the isosurface including the validation samples (light blue circles) that were suggested and grown using BO. A comparison of Figure 5b and 5c shows that the BO exploitation campaign led diamond growth and processing into a previously unexplored region of the parameter space. Importantly, we find that the reactor nitrogen concentration and irradiation dose must be optimized together, and a small optimal range was identified (see Figure 5c). Furthermore, with this analysis we were able to study each of these data points in the optimized region in depth to find that these samples demonstrate both clear improvements in coherence time and charge conversion compared to the previous best samples. It is likely that this growth regime has fewer acceptor defects due to the observation of better charge state stability, especially when measured at higher laser powers.[16] A more detailed study quantifying the various types of defects and their concentrations in this optimized region compared to a standard growth recipe is currently underway.

**Conclusion**

We use supervised machine learning trained with a unique data set comprised of over 100 quantum diamond growths with corresponding optical and quantum property characterization to demonstrate predictive performance of a magnetic field sensitivity figure of merit (FOM). Importantly, we utilize Bayesian optimization to grow a quantum diamond in underexplored growth parameter space that demonstrates a 300% relative improvement compared to average irradiated samples and a 55% relative improvement in FOM compared to the previous best sample. Use of machine learning led to the extraction of the most critical processing parameters, namely electron irradiation dose, diamond seed depth (i.e., distance from the plasma ball), miscut angle, and nitrogen in the chamber, which account for ~75% of the total sensitivity FOM. Additionally, multidimensional analysis revealed a new processing window for nitrogen incorporation and electron irradiation dose found by Bayesian optimization. This represents an advancement in quantum diamond growth, where disconnected process parameters are quantitatively related to one another for the first time. The growth, characterization, and ML framework presented in this work can be applied to optimize other quantum-relevant FOM's for color centers in diamond and for various types of materials growth reactors. For example, metrics such as single photon purity and indistinguishability for quantum interconnects could also be targeted. We envision this framework being applied to accelerate other solid-state quantum materials including transition metal dichalcogenides, hexagonal boron nitride, silicon carbide, inorganic complexes, and defects in other chemical systems.

## Methods

### Supervised Machine Learning Approach and Bayesian Optimization

All processing of the data, as well as the application of regressors is done in Python. The data inputs (X) are standardized using *StandardScaler* from scikit-learn, which scales the inputs to unit variance. Nine different machine-learning models are used: linear regression (LR), lasso, elastic net (EN), support-vector regression (SVR), random forest (RF), stacked regression of linear regression, support-vector regression, random forest, and ridge (ENS), stacked regression of ridge, lasso, and random forest using GridSearchCV (GridSearch), nearest neighbor (KNN), and gradient boost (GB), all from scikit-learn. For each model, regression is performed on the standardized and non-standardized data sets and compared using 5-fold cross-validated root mean square error (RMSE) of the test data.[32] Of these models, gradient boost is primarily used, as it is more appropriate for continuous, small datasets and furthermore, had the lowest cross-validated RMSE of the standardized dataset.[32,33] Additionally, for each model, the SHAP (Shapley Additive exPlanations) values are calculated, which quantify each feature's impact on the model prediction using the SHAP package. The PV-ML-Starter-Kit on GitHub is utilized to create a function that uses Bayesian optimization to offer combinations of features to maximize the output. This code uses the BayesianOptimization function from GPyOpt. Initially, the parameters of the Bayesian optimization function are tuned to be explorative, meaning the acquisition jitter was higher and the acquisition function used was the "lower confidence bound" (LCB). This means that the optimal solution regions that the function suggested were in low sampling areas and would improve the predictive power of the models. After 5 explorative validation growths, the acquisition jitter is lowered, and the acquisition function chosen was "expected improvement" (EI), meaning the Bayesian optimization is more tuned towards exploitation, or maximizing the output of machine learning. After changing the acquisition function and jitter, 12 samples are grown in order to maximize the FOM. In total, 18 diamonds are grown, resulting in 42 unique data points after post-processing to serve as validation growths.

### Optical Characterization and Quantum Measurements

### Photoluminescence Measurements

A home-built fluorescence microscope (532 nm laser) is used for all photoluminescence (PL)-based measurements. PL measurements are taken over a range of laser powers, from 50 uW to 50 mW. The fluorescence fraction from each charge state is determined through a deconvolution analysis, where each spectra is a linear combination of $NV^0$ and $NV^-$ spectral contributions. That is, for a normalized $NV^-$ basis function $I_-(\lambda)$ and normalized $NV^0$ basis function $I_0(\lambda)$, each acquired PL spectrum $I$ is fit to a linear combination of the basis functions $I(\lambda) = c_- I_-(\lambda) + c_0 I_0(\lambda)$, where $c_-$ and $c_0$ are the weights of each basis function whose ratio yields the desired charge state ratio.[25]

### Coherence Time Measurement Procedure

For coherence time measurements, the diamond is mounted such that the magnet is roughly aligned along an NV axis. This arrangement is preferred due to the (100) face and overall geometry of the diamond. The MW loop is positioned on top of the diamond with the focused laser spot in the center of the antenna loop. The diamond mount is adjusted to focus the confocal spot near the top of the diamond where the MW loop is located. This allows us to optically address and detect only the NVs that are near the antenna loop and that will experience a strong MW drive field.

A static field is applied with a projection $B\|$ along the NV axis, then to zeroth order the $|\pm 1\rangle$ states are split by $2\gamma_{NV}B\|$, where $\gamma_{NV}$ is the NV gyromagnetic ratio (2.8 MHz/G). CW-ODMR is then recorded where resonances occur at $E = D \pm \gamma eB\|$, where D is the zero field splitting (ZFS). With the static magnetic field aligned to one NV class, the microwaves address only one NV orientation. To implement the measurement protocols for $T_2^*$ and $T_2$, both the MW resonance frequency at which the fluorescence contrast is maximized (to maximize signal-to-noise) and the MW pulse durations that evolve the NV spin state to specific orientations are determined.

The pulse sequences used to measure the coherence time scales require different MW control pulses, which are calibrated for duration. For measuring $T_2$, for example, a MW $\pi$ pulse is needed to rotate the NV spin from $|0\rangle$ to $|^-1\rangle$. To measure the duration of this $\pi$ pulse (and other control pulses for different measurement protocols), the MW power is fixed, and the duration of an applied MW pulse is swept. The fluorescence contrast, defined for this pulse sequence as $m_1/m_0$, is then plotted against the MW pulse duration. The signal is damped due to inhomogeneous broadening from coupling to the environment and from gradients in the MW field. The shape of the curve is modelled as a sinusoidal function with a stretched exponential envelope. To measure coherence times, there are three required pulse times: a $\pi/2$ pulse that creates an equal superposition of the $|0\rangle$ and $|^-1\rangle$ spin states, a $\pi$ pulse that inverts the spin state from $|0\rangle$ to $|^-1\rangle$, and a $3\pi/2$ pulse that creates another equal superposition state but with a different relative phase. These times correspond to when the cosine has a phase of $\pi/2$, $\pi$, and $3\pi/2$ and the times are extracted from the aforementioned fit.

**Author Contributions**

D.W.D., E.P., L.P., and D.B. conceived and designed the experiments and analysis. D.W.D., E.P., J.M., A.M., and T.O. prepared and characterized all diamond seeds and grew the quantum diamond samples using PE-CVD. E.P. performed the optical characterization of the quantum diamond materials with support from D.W.D. and L.P. A.K. wrote the code for the database with input from D.W.D., E.P., L.P., J.M., A.M., and T.O.. J.S., L.P., D.W.D., J.M., E.P., and D.B. developed the equation for the magnetic field sensitivity figure of merit. S.V., E.P., and D.W.D developed the machine learning and Bayesian optimization code to incorporate the database features and outputs. G.W., B.L. performed quantum measurements on diamond samples with supervision from P.C.. D.W.D. and E.P. wrote the first draft of the manuscript with the first set of edits by C.M., then all authors contributed feedback and comments. D.W.D, J.M., P.C., E.P., and D.B. directed and supervised the research.

**Acknowledgements**

D.W.D, E.P., L.P., J.M., A.M., T.O., and D.B. acknowledge support for this project from the Air Force Research Laboratories (AFRL) under award No. FA8702-15-D-0001. G.W., B.L., and P.C. acknowledge support from DARPA DRINQS program (Cooperative Agreement No. D18AC00024). G.W. thanks MathWorks for their support in the form of a Graduate Student Fellowship. The opinions and views expressed in this publication are from the authors and not necessarily from MathWorks. D.W.D. and E.P. thank Krithika Manohar (U. Washington), Tonio Buonassisi (MIT), Aleks Siemens (MIT), Lin Li (MIT-LL), and Armi Tiihonen (MIT/Alto University) for valuable discussions related to machine learning.

## Competing Interests

The authors declare no competing interests.

## Data Availability

The data that support the findings of this study are available, under certain restrictions, from the corresponding authors upon reasonable request.

## Code Availability

The MATLAB and Python code used in this work is limited to sharing under restrictions from funding agencies.

**Supplementary Information for:**

**Machine Learning Enables Optimization of Diamond for Quantum Applications**

Dane W. deQuilettes,[1,*] Eden Price,[1*] Linh M. Pham,[1] Arthur Kurlej,[1] Swaroop Vattam,[1] Alexander Melville,[1] Tom Osadchy,[1] Boning Li,[2] Guoqing Wang,[2,3] Collin N. Muniz,[1] Paola Cappellaro,[2,3] Jennifer M. Schloss,[1] Justin L. Mallek,[1] Danielle A. Braje[1]

1. Massachusetts Institute of Technology, Lincoln Laboratory, Lexington, MA, 02421, USA.

2. Department of Physics, Massachusetts Institute of Technology, Cambridge, MA, 02139, USA.

3. Department of Nuclear Science and Engineering, Massachusetts Institute of Technology, Cambridge, MA, 02139, USA.

**\*Equal Contribution**

**Corresponding Author:** braje@ll.mit.edu

## Diamond Seed Characterization

We use a benchtop X-ray diffraction system to measure the Laue diffraction patterns of single-crystal diamond seed samples before growth. For quantitative determination of seed miscut angles, the sample is loaded into a sample holder, with the axes defined in Supplementary Figure 1b. The tilt of the sample relative to the seed holder is determined using optical displacement measurements at the corners of each seed.

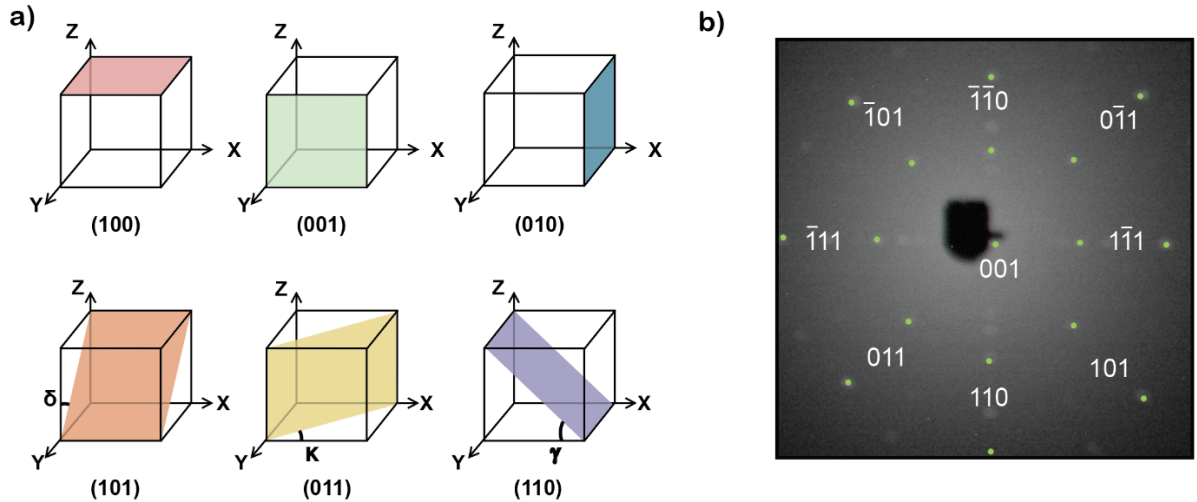

**Supplementary Figure 1.** a) Miller indices representation of the diamond crystal planes and how the three miscut angles delta, kappa, and gamma are defined relative to these planes. b) Laue diffraction pattern of a diamond seed before PE-CVD growth used to determine miscut angles with the measured diffraction peaks (faint white circles) and the known experimental diffraction patterns overlayed (green dots) with the corresponding Miller indices.

Next, we perform the X-ray measurement and collect four 40-second frames of the diamond's X-ray diffraction patterns to determine how far from the crystallographic axes the diamond has been cut. To determine the miscuts from these axes, the known experimental diffraction pattern of diamond with no miscut angle is overlayed on the collected diffraction pattern. These known axes are manually adjusted so that the known crystallographic axes are aligned to the corresponding measured axes in the X-ray images in order to see how far the measured axes are miscut from the known axes (see Supplementary Figure 1b). The miscut angle kappa ($\kappa$) represents the miscut angle from the (001) plane (or rotation of X-Y plane) axis, while miscut angles gamma ($\gamma$) and delta ($\delta$) represent the angles from the (110) and the (101) planes respectively.

Because the diamond is symmetric about the (100), the axes definitions of X and Y are interchangeable when the seed is removed from the diamond holder. Therefore, we simplify the angles of $\gamma$ and $\delta$ so that they represent total miscut normal to the (100) face. In order to calculate this total miscut, called miscut theta ($\theta$) or simply "miscut angle", we use the following formula described elsewhere[1]:

$$\theta \,(\deg°) = \sqrt{|\gamma|^2 + |\delta|^2}$$

*(Eq. 1)*

Which represents the angle to the normal vector to the surface (100 plane) of the diamond. These values are measured for every seed in this study.

**Database Structure and Implementation**

The backend of the database is implemented with MySQL and the front-end implemented in MATLAB. MySQL is a relational database, which is a framework that naturally maps to the diamond growth workflow we have described in the main article. In a relational database, information is grouped into tables (where a row identifies an instance of information and the columns identify features of that instance) and these tables are then related to each-other through a common "key" denoting a relationship between a row of information in one table to another.

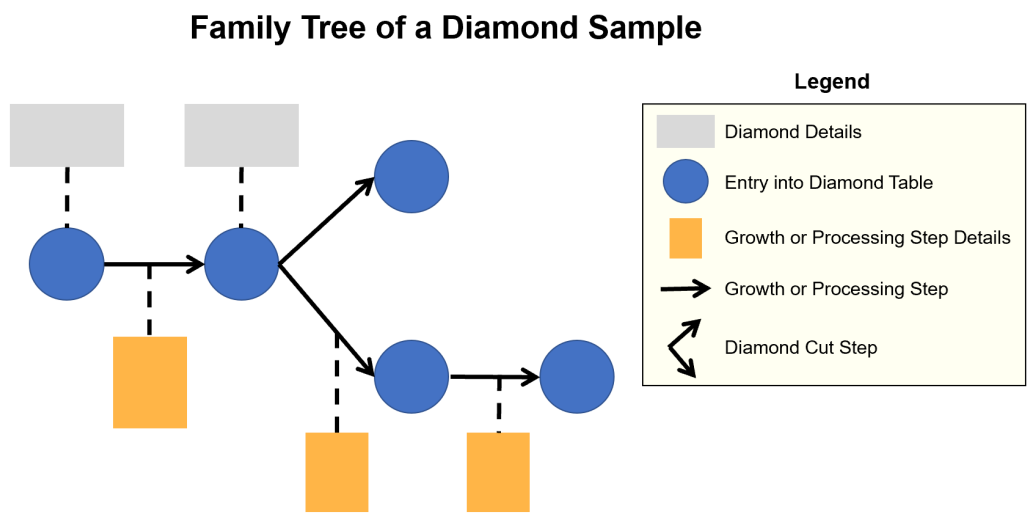

**Supplementary Figure 2.** Schematic representing the 'family tree' of a diamond sample. Blue circles represent an entry into a diamond table (i.e., parent diamond, child diamond). Solid black arrows represent entries into a diamond processing table (i.e., parent diamond, child diamond, CVD growth, processing step electron irradiation, dicing, etc.). Horizontal gray rectangles represent entries into other tables that describe the diamond at that step (i.e., quantum measurement results, location, images, etc.). Vertical orange rectangles represent entries into other tables that describe further describe the processing done at that step (i.e., growth properties, electron irradiation dose, dicing information, etc.). Multiple such trees exist within the database for each new 'lineage' of diamonds.

The processing and characterization history of a diamond can be tracked as a graph somewhat akin to a family tree as shown in Supplemental Figure 2. Vertices in the tree represent a snapshot of a diamond in this step of processing while edges represent the process step (a diamond was processed and yielded a 'new' diamond). The 'vertex' and the 'edge' are each described with a MySQL table (representing the diamond sample and the diamond processing step), and additional MySQL tables can be introduced and related to either the 'vertex' or 'edge' to further describe either the sample (characterization data, location data, images, etc.) or the processing step (growth, anneal, irradiation parameters, etc.).

The database is semi-automatically populated, through users entering information related to the diamond, process step, or characterization using a MATLAB GUI or by the MySQL server periodically searching structured directories for appropriate information. This same MATLAB GUI can also be used to present a snapshot of the status of the tracked samples or can export data into a variety of file types where further analysis can be performed.

## NV⁻ (ppb) Estimate using Cryo UV-Vis

In order to get an estimation of the concentration of NV⁻, we conducted several measurements of the absorption features at 637 nm in a cryogenic UV-Vis set-up. We examined electron-irradiated and annealed samples in the Renishaw Qontor Confocal Raman Microscope by illuminating the samples with a calibrated white light source. We used a Linkham LNP96 cryostage system to reach temperatures of 77 K in ~90 seconds. Using a 1200 l/mm grating, we collect wavelengths from ~630-640nm for a total collection time of 200 ms. Using custom software, we convert the transmission to absorption coefficient and integrate the area under the NV⁻ peak (Supplementary Figure 3), and then convert the integrated absorption coefficient to NV⁻ (ppm) using the NV⁻ calibration constant.[2] To reduce the amount of time required to measure each sample, we generated a calibration plot of the integrated NV⁻ PL intensity (deconvolved from the basis function analysis) versus the NV⁻ density determined from cryo UV-Vis using a set of 8 samples with a large range in NV⁻ densities. This calibration plot was verified using a separate XY8 measurement to determine the NV⁻ density. We then use the calibration plot to convert the integrated NV- PL intensity to NV⁻ (ppb).

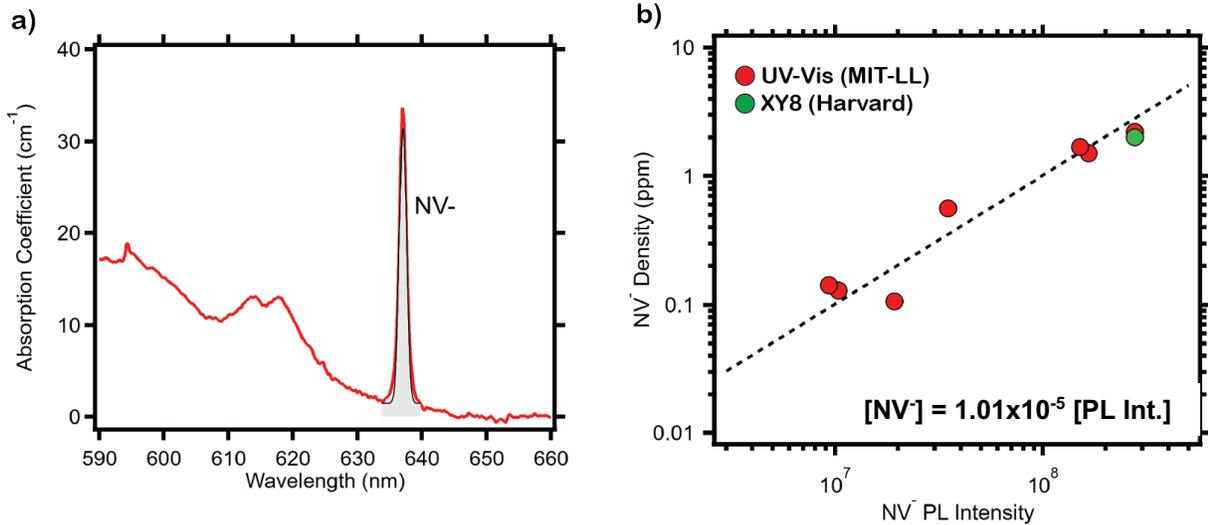

**Supplementary Figure 3.** a) Cryo (77 K) UV-Vis absorption coefficient spectrum of N-doped diamond. The peak at 637 nm is integrated to determine the NV⁻ density. b) NV⁻ density versus integrated NV⁻ PL intensity calibration plot used in order to determine the concentration of NV⁻ in all samples reported in this study.

## Sensitivity Figure of Merit (FOM)

The Sensitivity Figure of Merit described in this paper is a proxy for magnetometer sensitivity and is used as the target output for supervised machine learning. We begin with the equation for the shot-noise limited sensitivity for a Ramsey scheme using and ensemble of quantum sensors as shown in equation 1 and described elsewhere[1],

$$\eta \sim \frac{1}{g_e \mu_B} \frac{1}{C} \frac{1}{\sqrt{\beta n_{NV^-} V}} \frac{\sqrt{T_{dead} + T_2^*}}{T_2^*}$$

<div style="text-align:right">(Eq. 2)</div>

Here $g_e$ is the gyromagnetic ratio, $\mu_B$ is the Bohr magneton, $C$ is the measurement contrast, $\beta$ is the collection efficiency, $V$ is the excitation volume, $n_{NV-}$ is the NV density, $T_{dead}$ is the dead time for initialization and readout, and $T_2^*$ is the Ramsey coherence time.

We assume that

$$\eta \propto \frac{\sqrt{T_{dead} + \tau}}{C\sqrt{N}\,\tau}$$

*(Eq. 2)*

Where $C$ is contrast, $N$ is number of photons detected per measurement, $\tau$ is sensing duration, and $T_{dead}$ is dead time required for initialization and readout. More specifically,

$$T_{dead} = 1\mu s * \frac{s+1}{s}$$

*(Eq. 3)*

Where $s = \frac{I}{I_{sat}}$, $I$ is the laser intensity, $I_{sat}$ is the saturation intensity ($\sim 3\text{mW}/\mu m^2$) and

$$C = C_{NV-} * \frac{PL_{NV-}(s)}{PL_{NV-}(s) + PL_{NV0}(s) + PL_{SiV-}(s)},$$

*(Eq. 4)*

Which is the power-dependent PL charge state efficiency times a constant. $N$ is set to be constant for all laser powers. For optimized sensing, dead time is increased until the same number of photons are detected per measurement. In order to compare across diamonds, we choose this constant $C_{NV-}$ to be the concentration of NV⁻ in parts per billion. This value is equal to

$$C_{NV-}(ppb) = PL_{NV-}(s_0) * 8.875E^{-5}$$

*(Eq. 5)*

Where $PL_{NV-}(s_0)$ is the contribution to PL from NV⁻ at a fixed laser power, and $8.875E^{-5}$ is a calibration factor from cryo UV-Vis measurements (see Supplementary Information Section X).

For magnetometry purposes, $\tau$ (eq. 2) is equal to $T_2^*$, but can be sent to other relevant spin lifetimes for different sensing protocols. Combining all of this information, the constants are dropped, making equation 2 become

$$\eta \propto \frac{\sqrt{T_2^* + 1\mu s * \frac{s+1}{s}}}{C_{NV-}(ppb) * \frac{PL_{NV-}(s)}{PL_{tot}(s)} * T_2^*}$$

*(Eq. 6)*

We choose the FOM to be proportional to $\frac{1}{\eta}$, so that

$$FOM = C_{NV^-}(ppb) * \frac{PL_{NV^-}(s)}{PL_{tot}(s)} * \frac{T_2^*}{\sqrt{T_2^* + 1\mu s * \frac{s+1}{s}}}$$

*(Eq. 7)*